\documentclass[conference]{IEEEtran}
\usepackage{graphicx}
\usepackage{amsmath}
\usepackage{amsfonts}
\usepackage{hyperref}
\usepackage{cite}
\usepackage{tocloft}
\title{MindCraft: Revolutionizing Education through AI-Powered Personalized Learning and Mentorship for Rural India}

\author{\IEEEauthorblockN{1\textsuperscript{st} Arihant Bardia}
\IEEEauthorblockA{\textit{Department of Information Technology} \\
\textit{Pune Institute of Computer Technology}\\
Pune, India \\
arihant.bardia123@gmail.com}
\and
\IEEEauthorblockN{2\textsuperscript{nd} Aayush Agrawal}
\IEEEauthorblockA{\textit{Department of Information Technology} \\
\textit{Pune Institute of Computer Technology}\\
Pune, India \\
asagrawal612@gmail.com}}
\begin{document}

\maketitle
%%%%%%%%%%%ABSTRACT%%%%%%%%%%%%%
\begin{abstract}
MindCraft is a modern platform designed to revolutionize education in rural India by leveraging Artificial Intelligence (AI) to create personalized learning experiences, provide mentorship, and foster resource-sharing. In a country where access to quality education is deeply influenced by geography and socio-economic status, rural students often face significant barriers in their educational journeys. MindCraft aims to bridge this gap by utilizing AI to create tailored learning paths, connect students with mentors, and enable a collaborative network of educational resources that transcends both physical and digital divides.

This paper explores the challenges faced by rural students, the transformative potential of AI, and how MindCraft offers a scalable, sustainable solution for an equitable education system. By focusing on inclusivity, personalized learning, and mentorship, MindCraft seeks to empower rural students, equipping them with the skills, knowledge, and opportunities needed to thrive in an increasingly digital world. Ultimately, MindCraft envisions a future in which technology not only bridges educational gaps but also becomes the driving force for a more inclusive and empowered society.
\end{abstract}


%%%%%%%%%INTRODUCTION%%%%%%%%%%%%%
\section{Introduction}
Education in India is marred with the aid of extensive disparities between rural and concrete regions. Regardless of technological advancements, rural college students face boundaries together with insufficient infrastructure, confined get right of entry to to educational sources, and the absence of mentorship. At the same time as city college students advantage from international-class schooling structures, those in rural India continue to be disadvantaged of possibilities that would empower them to triumph over socio-financial challenges. The virtual divide is one of the number one barriers, with many college students unable to get admission to digital studying tools due to a lack of devices and net connectivity.

MindCraft, a generation-driven academic platform, aims to bridge this gap with the aid of leveraging artificial Intelligence (AI) to create personalized learning pathways for rural college students. Via AI, MindCraft gives tailored academic content, mentorship, and collaborative sources, ensuring that scholars get hold of a complete learning revel in. The platform’s recognition on inclusivity and mentorship is designed to equip rural college students with the skills vital to thrive in an increasingly virtual global.

The foundation in the back of MindCraft comes from a deeply personal experience with a local tea vendor near my university. Regardless of his financial struggles, he desires of offering higher instructional opportunities for his daughter, a category 8 student. This encounter highlighted the want for scalable, generation-pushed solutions to uplift students from rural regions. MindCraft represents a attention of that imaginative and prescient, aiming to empower the subsequent era of students from small cities and villages.
%%%%%%%%%BACKGROUND AND MOTIVATION%%%%%%%%%%%%%%%%
\section{Background and Motivation}
The academic disparity between rural and urban India is a significant challenge, as many rural college students struggle to access even the most basic learning resources. This gap goes beyond mere physical infrastructure deficiencies; it also encompasses the absence of mentorship and career guidance, both of which are critical in shaping brighter, more promising futures. MindCraft was founded with a vision to disrupt this cycle by providing a platform that offers tailored learning experiences and a mentorship network to address these very gaps. The goal is not only to bridge the resource gap but to offer a comprehensive ecosystem that integrates AI, mentorship, and resource-sharing.

By utilizing AI technology, MindCraft can deliver personalized learning journeys that cater to each student’s unique needs, ensuring that the educational experience is both engaging and effective. Furthermore, the platform serves as a mentorship hub, connecting students with experienced professionals who can provide guidance, career advice, and support. This dual approach of using technology for learning and human connection for growth fosters an environment where both students and educators can thrive. Ultimately, MindCraft aims to empower students by equipping them with the tools, knowledge, and mentorship required to succeed, while simultaneously offering teachers the resources and support they need to deliver high-quality education in remote and underserved areas.

%%%%%%%%%%%%%%%%%%%%%%%%%%%%%%%%%%%%%%%%%%
\section{Related Work}
The developing adoption of synthetic intelligence (AI) in education has spurred improvements in customized learning and career guidance systems. Our MindCraft mission, which integrates customized education, mentorship, and career counseling, builds on numerous foundational studies in these areas.

Murtaza et al. \cite{ref1} survey AI-based personalized learning, highlighting challenges like scalability, data privacy, and the digital divide—key concerns for equitable and adaptive education. Prior research explores AI-driven recommendation systems, adaptive learning frameworks, and real-time analytics, yet challenges remain in scalability and accessibility. Our work builds on these studies by reviewing existing solutions and proposing an AI-driven framework to enhance personalized learning while addressing these limitations.

Laak and Aru \cite{ref2} analyze AI-based Personalized learning solutions through the lens of the OECD Learning Compass 2030, identifying gaps in collaboration, cognitive engagement, and competency development. They emphasize that while current PL technologies aid learning, true personalization requires a holistic transformation of the educational system. Building on these insights, our work explores AI-driven frameworks that enhance personalization while incorporating essential elements of modern education.

Within career development, Ghuge et al. \cite{ref3} illustrate the effectiveness of data-driven AI in delivering customized career counseling, a principle we incorporate in our AI-powered career module. Their system assesses students' academic profiles, extracurricular activities, and past performance to generate personalized career recommendations, reducing academic misalignment and improving decision-making. While existing AI-based approaches provide valuable insights, integrating them into a holistic educational framework remains a challenge. Our work builds on these foundations, aiming to offer a more adaptive and comprehensive AI-driven learning and mentorship system.

AI-driven personalized learning systems leverage recommendation technologies to enhance online education. Gm et al. \cite{ref4} review various recommendation approaches, including collaborative and content-based filtering, and highlight a shift toward machine learning for personalization. However, they identify challenges such as content misinterpretation, student disengagement, language barriers, and infrastructure limitations. Their work suggests integrating emerging technologies like Fluxy AI and AI-powered virtual proctoring to create more dynamic learning environments. Building on these insights, our research explores AI-driven personalized learning frameworks that address these challenges while ensuring adaptability, accessibility, and scalability in education.

The Indian education system faces significant challenges in adapting to modern learning approaches. Shinde et al. \cite{ref5} review the current state of education in India, emphasizing the need for systemic upgrades to keep pace with globalization. Their study highlights issues in technical education, including admission conditions and student performance, using the fishbone diagram technique to analyze root causes of failure. While their work focuses on structural challenges, our research builds upon these insights by proposing AI-driven personalized learning frameworks to enhance accessibility, adaptability, and student engagement in the evolving Indian education landscape.

Collaborative learning plays a crucial role in personalized education, and effective group formation is essential for maximizing learning outcomes. Ramos et al. \cite{ref6} propose a novel genetic algorithm for optimizing group formation in collaborative learning environments within Learning Management Systems (LMSs). Their study highlights how innovative genetic operators, such as modified crossover techniques, enhance accuracy and efficiency in forming balanced learning groups. While their work focuses on algorithmic improvements, our research integrates AI-driven methodologies to personalize learning beyond group formation, incorporating mentorship, adaptive content delivery, and real-time feedback mechanisms to enhance student engagement and collaboration.

AI-powered mentorship platforms have the potential to transform professional and educational guidance. Bagai and Mane \cite{ref7} explore the conceptual design of MentorAI, an AI-driven mentorship platform aimed at career progression, skill development, and personalized guidance. Their study highlights the key technological components, including AI, machine learning, and natural language processing, which enable real-time, context-sensitive mentorship. However, they also acknowledge challenges such as data privacy, security, and algorithmic bias. Building on these insights, our research integrates AI-driven mentorship into personalized learning, ensuring adaptive guidance while addressing ethical concerns and the human-AI balance in educational mentorship.

AI-driven mentorship platforms are increasingly being explored to enhance career guidance and skill development. Vidhya et al. \cite{ref8} propose a Career Mentorship Platform that leverages AI-powered recommendations to optimize mentor-mentee matching, facilitate real-time interactions via chatbots, and integrate scheduling and feedback mechanisms. Their work emphasizes the role of AI in streamlining mentorship processes. Building on this, our research integrates AI-driven mentorship within personalized learning, ensuring adaptive career guidance that aligns with students’ educational progress and evolving professional aspirations.

Tewari \cite{ref9} provides a panoramic view of 150 years of higher education in India, highlighting key developments, policy changes, and systemic challenges. While traditional university education has played a vital role in shaping academic frameworks, modern AI-driven educational platforms are now emerging to bridge existing gaps, particularly in accessibility and personalization. Our research builds on this evolution, leveraging AI to create personalized learning and mentorship experiences that address contemporary educational challenges in India.

Chimalakonda \cite{ref10} introduces GAMBLE, a goal-driven model-based learning environment aimed at improving the quality of education. The framework emphasizes explicitly defining learning objectives and refining instructional models to achieve these goals. Additionally, the study proposes SPLEAN, which integrates lean thinking with software product lines to enhance the productivity of e-learning systems. These methodologies provide a structured approach to designing adaptive learning environments, aligning with our vision of AI-powered personalized education. By leveraging similar goal-driven strategies, our platform seeks to ensure quality education while addressing scalability and accessibility challenges in rural India..

Zimmerman \cite{ref11} presents a foundational framework for Self-Regulated Learning (SRL), emphasizing learners' ability to actively participate in their educational journey by setting goals, monitoring progress, and adjusting strategies. The study highlights cognitive, metacognitive, and motivational aspects that influence learning outcomes. This aligns with our approach in MindCraft, where AI-powered personalized learning paths are designed to foster self-regulated learning. By integrating SRL principles into adaptive learning models, our platform aims to empower students to take control of their educational experiences, bridging the gap between guidance and independent learning.

Vandewaetere and Clarebout \cite{ref12} provide a comprehensive overview of advanced technologies in personalized learning, emphasizing the role of learner models in dynamically adapting instruction based on cognitive and noncognitive needs. Their study highlights the integration of Artificial Intelligence (AI) and Educational Data Mining (EDM) in shaping intelligent tutoring systems and refining learner behavior analysis. This aligns with MindCraft's vision of AI-driven personalized learning, where real-time learner modeling ensures adaptive learning experiences. By leveraging AI-powered mentorship and adaptive content delivery, MindCraft aims to enhance engagement, self-regulated learning, and academic performance, bridging the urban-rural educational divide.

Together, these studies provide a comprehensive foundation for MindCraft, which unifies personalized learning, dynamic content delivery, and career counseling into a single platform to address the multifaceted challenges of current schooling.

%%%%%%%%%%%%%%%%%%%%%%%%%%%%%%%%%%%%%%%%%%%%%%
\section{Challenges in Rural Education and AI Adoption}

India’s rural education system faces numerous challenges, including a lack of infrastructure, teacher shortages, and limited access to educational resources. These challenges are compounded by a lack of digital literacy and poor internet connectivity, which hinders the adoption of AI-powered learning solutions.

\subsection{The Digital Divide}
The digital divide remains a significant barrier to the significant adoption of era in training. Many rural regions lack reliable internet connectivity and get right of entry to to gadgets which include computers or smartphones. As a end result, rural college students often miss out on the blessings of on line training platforms, making it difficult for them to compete with their urban opposite numbers.

MindCraft objectives to deal with this divide with the aid of providing low-bandwidth alternatives for gaining access to content and ensuring compatibility with low-fee devices. Additionally, partnerships with nearby governments and agencies can assist provide the essential infrastructure to facilitate great adoption.

\subsection{Language Barriers}
Another mission in rural training is the language barrier. Many instructional assets are to be had in English, which might not be the number one language for rural college students. MindCraft addresses this trouble via supplying content material in more than one local languages, making it reachable to a larger population and ensuring that scholars can examine of their desired language.

\subsection{Teacher Training and Support}
The successful implementation of AI in education also depends on the preparedness of instructors to integrate technology into their classrooms. While many instructors in urban areas are trained to use digital tools, rural educators often lack the essential skills and resources. MindCraft addresses this issue by providing professional development opportunities for instructors, empowering them to use AI tools effectively in their teaching.

\subsection{Geographical Isolation}
Rural students often face the additional challenge of geographical isolation, which exacerbates the academic divide. Many rural areas are located in remote or hard-to-reach regions, making it hard for students to get access to educational institutions and resources. In some regions, the closest school or learning center can be several kilometers away, and students are often forced to travel long distances on foot or by unreliable transportation. This geographical isolation limits students' ability to participate in extracurricular activities, interact with peers, and receive adequate support from instructors and mentors.

MindCraft overcomes this challenge by providing a fully digital platform that can be accessed remotely, allowing students from even the most isolated regions to experience the same education and mentorship opportunities as their urban counterparts. The platform ensures that rural students can access educational content and engage with mentors and peers without the need for physical proximity.

%%%%%%%%%%%%%%%%%%%%%%%%%%%%%%%%%%%%%%%%%%%%%%%%%%
\section{Case Study: Empowering Rural Education – Ravi’s Journey with MindCraft}

\subsection{Background: The Struggles of Rural Education}
Ravi, a 14-year-old scholar from a remote village in Madhya Pradesh, has always been eager to learn, but his circumstances hold him back. With only two instructors for over 150 students, personalized attention is a luxury he can't afford. He struggles with mathematics and English, leading to a lack of self-confidence. Most online learning resources are in English, making self-study hard. No career guidance, mentorship, or exposure to technology means he has no clear vision for his future.

In spite of his potential, Ravi’s educational journey seems predetermined—until he is introduced to MindCraft, an AI-powered education and mentorship platform designed to bridge the gap in India’s rural education system.

\subsection{How MindCraft Transforms Ravi’s Learning Experience}

\subsubsection{AI-Powered Personalized Learning: Breaking the One-Size-Fits-All Barrier}
\begin{itemize}
\item Upon logging into MindCraft, Ravi takes a skillful evaluation, which identifies his weak areas in fractions, algebra, and English grammar.

\item The AI creates a personalized learning path, breaking down complicated subjects into bite-sized lessons tailored to his learning pace.

\item Lessons are provided in Hindi and English, ensuring language is no longer a barrier.

\item AI-generated quizzes dynamically adjust difficulty levels, reinforcing concepts without overwhelming him.

\end{itemize}

\subsubsection{Real-Time AI Tutor: Learning Without Waiting}
\begin{itemize}
   \item While Ravi struggles with a concept, MindCraft’s AI guide provides a step-by-step explanation in simple terms, with visual aids and interactive problem-solving.

    \item Unlike a crowded classroom where his doubts go unheard, he can ask unlimited questions without hesitation.

    \item With practice exercises and instant feedback, he progresses at his own pace, improving his accuracy and confidence.
\end{itemize}

\subsubsection{Career Discovery \& Mentorship: Expanding Ravi’s World}
\begin{itemize}
  \item For the first time, Ravi explores career paths beyond what he sees around him.

\item MindCraft’s AI-based career exploration tool suggests coding as a potential interest.

\item Through the platform’s mentorship program, he connects with an engineering student who guides him through basic programming concepts.

\item Excited by this new world, Ravi begins using MindCraft’s interactive coding module, where he builds a simple Python-based calculator—his first-ever software challenge.

\end{itemize}

\subsubsection{Overcoming Rural Limitations: Education Beyond Connectivity}
\begin{itemize}
   \item Since his village has limited internet access, MindCraft’s offline mode allows him to download lessons and quizzes for later use.

\item His school integrates AI-generated lesson plans, helping teachers deliver structured and engaging lessons without extra workload.

\item Peer-to-peer collaborative challenges encourage Ravi and his classmates to compete in gamified learning, making education fun and engaging
\end{itemize}

\subsection{The Transformation: Six Months Later}
Ravi’s journey with MindCraft creates a visible impact on his education and future aspirations:
\begin{itemize}
 \item  Mathematics and English, once his weak points, become his strengths, improving his exam scores by 40%.

\item He no longer skips school; rather, he actively participates in class, helping friends with problem-solving.

\item Exposure to coding and mentorship broadens his vision, and he now dreams of becoming a software engineer.

\item His teachers notice a change—Ravi is more confident, curious, and engaged in learning.
\end{itemize}

\subsection{Conclusion: The Power of AI in Rural Education}
Ravi’s case demonstrates how MindCraft’s AI-driven, personalized learning and mentorship model can break traditional barriers in rural education. With adaptive learning, real-time AI assistance, career exploration, and an accessibility-focused design, MindCraft has the potential to redefine learning for millions of students like Ravi, empowering them with knowledge, confidence, and a future full of opportunities.

By leveraging technology to personalize education, MindCraft isn't just helping students pass exams—it is creating a generation of learners who can dream beyond their circumstances.

%%%%%%%%%%%%%%%%%%%%%%%%%%%%%%%
\section{MindCraft Platform: Features and Functionality}
MindCraft is constructed on a robust technological stack that consists of AI, device learning, and facts analytics to supply personalized gaining knowledge of experiences and mentorship. Beneath are the key functions of the platform:
\begin{figure*}[ht]
    \centering
    \includegraphics[width=\textwidth, height=10cm]{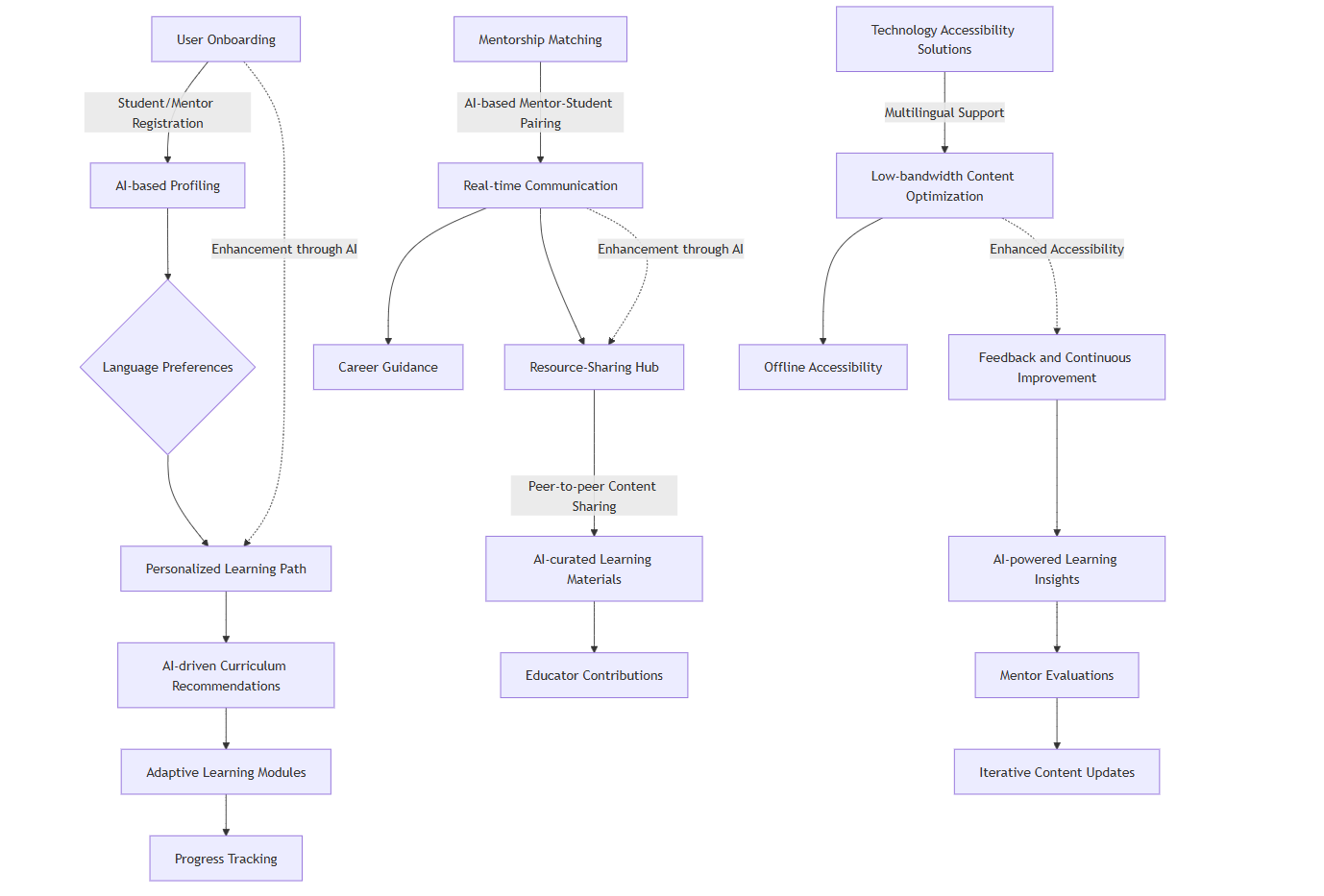} % Change filename accordingly
    \caption{Flowchart for MindCraft}
    \label{fig:flowchart}
\end{figure*}
\subsection{Personalised Gaining Knowledge Paths}
AI algorithms analyze every scholar’s performance and studying preferences, adapting the content material and difficulty stage to fit their wishes. This guarantees that every scholar gets a tailored studying experience, which improves retention and engagement.
\subsection{AI-Powered Mentorship}
The platform connects students with mentors who offer academic and profession guidance. The usage of AI, the platform fits students with mentors based totally on their interests, career aspirations, and academic overall performance, making sure a meaningful mentorship experience.
\subsection{Collaborative Aid Sharing}
MindCraft fosters a network-driven technique to mastering through permitting students, teachers, and mentors to percentage educational resources, which includes take a look at materials, notes, and mission ideas. This collaborative surroundings complements the learning experience and enables college students to get entry to a wealth of expertise.
\subsection{Actual-Time Communication Tools}
The platform includes features for actual-time communication among students and mentors, in addition to peer-to-peer discussions, ensuring that students acquire the support they need when they need it.
\begin{figure}[ht]
    \centering
    \includegraphics[width=\linewidth, height=15cm]{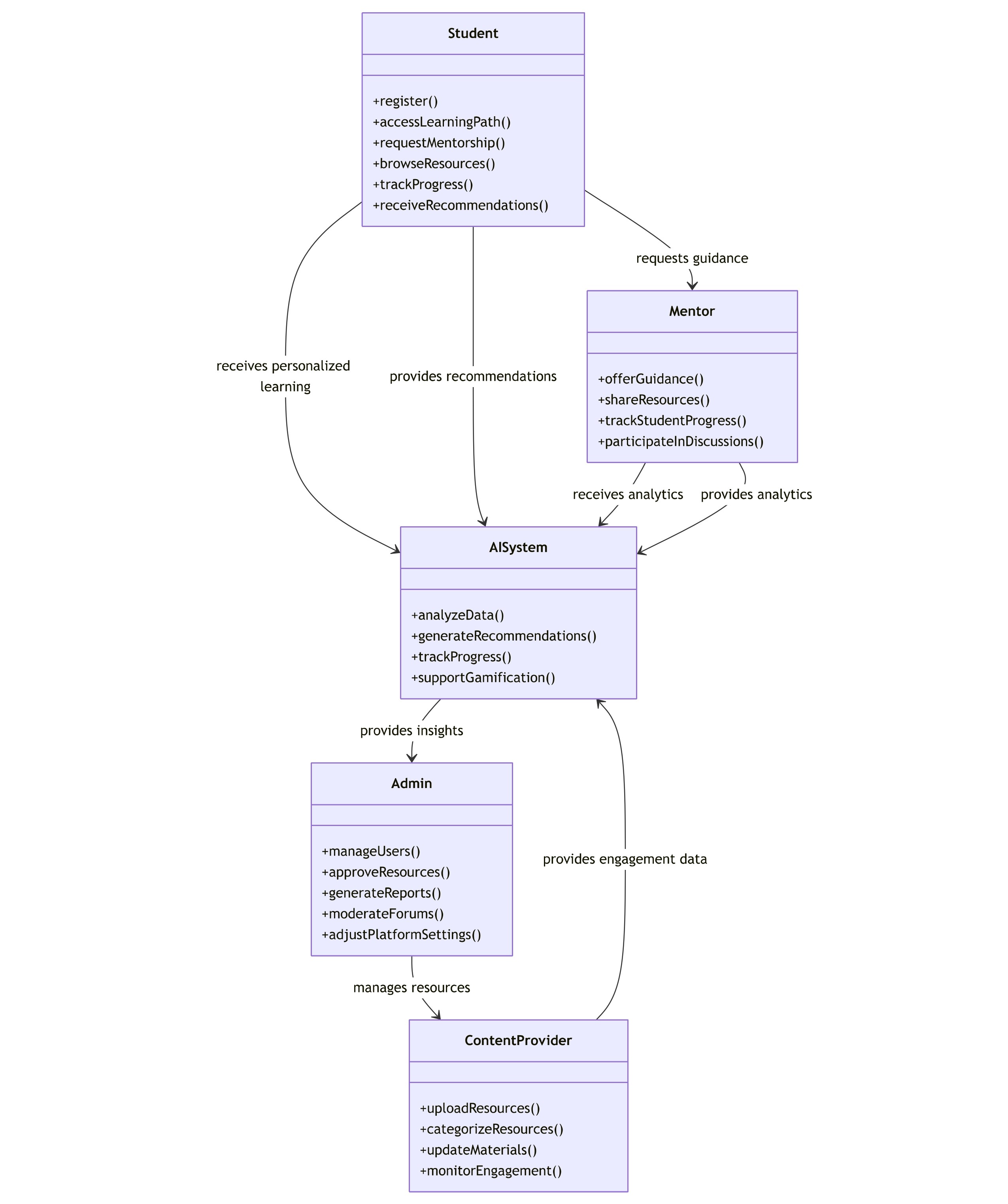} % Change filename accordingly
    \caption{Class Diagram for MindCraft}
    \label{fig:class_diagram}
\end{figure}

%%%%%%%%%%%%%%%%%%%%%%%%%%%%%%%%%%%%%%%%%
\section{Scalability and Sustainability}

\subsection{Scalability}
MindCraft has been designed to scale efficaciously, with the potential to feature new functions and enlarge to other areas. The platform's modular structure allows for the easy addition of new languages, courses, and mentorship areas as needed. This pliability ensures that MindCraft can develop alongside the evolving needs of rural students and the educational landscape.

\subsection{Investment and Partnerships}
To ensure long-term sustainability, we are exploring partnerships with educational institutions, NGOs, and corporate sponsors. Moreover, MindCraft plans to launch a crowdfunding campaign to secure funding for further development and outreach. These initiatives will aid MindCraft’s growth, expansion, and continuous improvement.

\section{Future Vision}

\subsection{Technological Advancements}
In the future, MindCraft aims to integrate AI-driven personalization further, allowing the platform to suggest learning paths based on each student's preferences, progress, and learning style. Additionally, we plan to explore virtual reality (VR) and augmented reality (AR) for hands-on learning experiences, further enhancing the interactivity and engagement of students.

\subsection{Global Expansion}
While MindCraft is currently focused on rural India, we see potential for expansion into other areas with similar educational challenges. The platform’s modular design makes it adaptable to different educational contexts and languages, allowing us to reach underserved communities worldwide.

\section{Platform Architecture and Design}

\subsection{ Technical Overview}
MindCraft is built with the goal of scalability, accessibility, and ease of use. The platform’s structure includes the following components:

\begin{itemize}
\item \textbf{Frontend:} The user interface is developed using React.js, offering a dynamic and responsive layout that adapts to different screen sizes and devices.
\item \textbf{Backend:} Node.js is used to manage the backend, ensuring fast and efficient data handling for mentorship, resources, and user interactions.
\item \textbf{Database:} MongoDB stores user information, resources, and mentor-mentee interactions in a scalable NoSQL format.
\item \textbf{Hosting:} The platform is deployed on Vercel for fast and reliable hosting, ensuring global accessibility.
\end{itemize}

\subsection{User Experience (UX)}
The platform’s layout prioritizes simplicity and ease of navigation. The aim is to create an intuitive user interface for students with limited technological experience. Key design considerations include:

\begin{itemize}
\item \textbf{Mobile Optimization:} Since many students in rural areas access the internet through mobile devices, the platform is fully optimized for mobile use.
\item \textbf{Low Data Consumption:} To accommodate users with limited internet connectivity, MindCraft’s design minimizes data usage while ensuring that key content is delivered efficiently.
\end{itemize}

\section{Impact of MindCraft on Rural Education}
\subsection{Improved Learning Outcomes}
Personalized learning has been proven to improve student outcomes by providing tailored content that meets the individual needs of each student. With MindCraft, students will benefit from an education designed specifically for their learning styles and pace.

\subsection{Access to Mentorship and Career Guidance}
Through the AI-powered mentorship feature, MindCraft connects students with experienced mentors who provide guidance on academic and career paths. This helps students make informed decisions about their future and fosters long-term success.

\subsection{Resource Availability and Sharing}
MindCraft enables a collaborative environment where students, teachers, and mentors can share resources. This collective knowledge-sharing significantly enhances the learning experience for all participants.
%%%%%%%%%%%%%%%%%%%%%%%%%%%%%%
\section{Conclusion}

MindCraft is a scalable, era-driven answer that ambitions to bridge the instructional hole in rural India. Via manner of leveraging AI, customized gaining of knowledge, and mentorship. The platform also gives students the gear and support they need to be first rate academically and professionally. Centering on inclusivity, collective effort, and sustainable practices, MindCraft has the aim for the transformation of rural schooling and to empower college students to break loose from socio-monetary limitations.

%%%%%%%%%%%%%%%%%%%%%%%%%%%%%%%%%%%%%%%%

\end{document}